\documentclass[10pt,aps,prl,superscriptaddress,amssymb,nobibnotes,reprint]{revtex4-1}

\usepackage[utf8x]{inputenc}
\usepackage{graphicx}
\usepackage{amsmath}
\usepackage{multirow}

\newcommand{\TNSi}{^{29}\mathrm{Si}}
\newcommand{\ndo}{\Downarrow}
\newcommand{\nup}{\Uparrow}
\newcommand{\edo}{\downarrow}
\newcommand{\eup}{\uparrow}

\renewcommand\figurename{Figure}  

\begin{document}
\title{A silicon quantum-dot-coupled nuclear spin qubit}
\author{Bas~Hensen}
\thanks{These authors contributed equally}
\author{Wister~Wei~Huang}
\thanks{These authors contributed equally}
\author{Chih-Hwan~Yang}
\author{Kok~Wai~Chan}
\author{Jun~Yoneda1}
\author{Tuomo~Tanttu}
\author{Fay~E.~Hudson}
\author{Arne~Laucht}
\affiliation{Centre for Quantum Computation and Communication Technology,
School of Electrical Engineering and Telecommunications,
The University of New South Wales, Sydney, New South Wales 2052, Australia}
\author{Kohei~M.~Itoh}
\affiliation{School of Fundamental Science and Technology, Keio University, 3-14-1 Hiyoshi, Kohoku-ku, 
Yokohama 223-8522, Japan}
\author{Thaddeus~D.~Ladd}
\affiliation{HRL Laboratories, LLC, 3011 Malibu Canyon Rd., Malibu, CA, 90265, USA}
\author{Andrea~Morello}
\author{Andrew~S.~Dzurak}
\email{b.hensen@unsw.edu.au,wister.huang@unsw.edu.au, a.dzurak@unsw.edu.au}
\affiliation{Centre for Quantum Computation and Communication Technology,
School of Electrical Engineering and Telecommunications,
The University of New South Wales, Sydney, New South Wales 2052, Australia}

\maketitle

\textbf{Single nuclear spins in the solid state have long been envisaged as a platform for quantum computing\cite{Kane1998silicon-based,Ladd2002All-Silicon,Skinner2003Hydrogenic}, due to their long coherence times\cite{Maurer2012Room-Temperature, Saeedi2013Room-Temperature,Muhonen2014Storing} and excellent controllability\cite{Vandersypen2005NMR}. Measurements can be performed via localised electrons, for example those in single atom dopants\cite{Pla2013High-fidelity,Pla2014Coherent} or crystal defects\cite{Childress2006Coherent,Neumann2010Single-Shot,Abobeih2018One-second}. However, establishing long-range interactions between multiple dopants or defects is challenging\cite{Bernien2013Heralded,Tosi2017Silicon}. Conversely, in lithographically-defined quantum dots, tuneable interdot electron tunnelling allows direct coupling of electron spin-based qubits in neighbouring dots\cite{Nowack2011Single-Shot,Veldhorst2014addressable,Veldhorst2015two-qubit,Watson2018programmable,Zajac2018Resonantly,Huang2019Fidelity}. Moreover, compatibility with semiconductor fabrication techniques\cite{Maurand2016CMOS} provides a compelling route to scaling to large numbers of qubits. Unfortunately, hyperfine interactions are typically too weak to address single nuclei. Here we show that for electrons in silicon metal–oxide–semiconductor quantum dots the hyperfine interaction is sufficient to initialise, read-out and control single silicon-29 nuclear spins, yielding a combination of the long coherence times of nuclear spins with the flexibility and scalability of quantum dot systems. We demonstrate high-fidelity projective readout and control of the nuclear spin qubit, as well as entanglement between the nuclear and electron spins. Crucially, we find that both the nuclear spin and electron spin retain their coherence while moving the electron between quantum dots, paving the way to long range nuclear-nuclear entanglement via electron shuttling\cite{Skinner2003Hydrogenic}. Our results establish nuclear spins in quantum dots as a powerful new resource for quantum processing.}

Electrons bound to single dopant atoms or localised crystal defects strongly interact with the host donor or defect nuclear spins\cite{Pla2013High-fidelity,Neumann2010Single-Shot}, as well as nearby lattice nuclear spins\cite{Pla2014Coherent,Childress2006Coherent}, due to their highly confined wavefunctions. In quantum dots, the electron wavefunction is less confined and typically overlaps with many more nuclear spins, leading to undesired effects such as loss of coherence and spin relaxation\cite{Johnson2005Tripletsinglet,Chekhovich2013Nuclear}. In silicon metal–oxide–semiconductor quantum dots, however, the strong confinement of the electrons against the $\mathrm{Si-SiO_2}$ interface, together with the possibility of small gate dimensions, result in a relatively small electron wavefunction\cite{Yang2013Spin-valley}, see Fig.~1a,b. This leads to strong hyperfine interactions, and when using isotopically enriched $^{28}\mathrm{Si}$ base material, the number of interacting $\TNSi$ nuclei may be only a few. Simulations of the distribution of expected hyperfine couplings\cite{Assali2011Hyperfine} in a quantum dot with a 8~nm wavefunction diameter and 800~ppm $\TNSi$ nuclei indicate an expectation of two to three $\TNSi$ nuclei per quantum dot that have a resolvable hyperfine coupling ($\geq 100$~kHz), and a maximally possible hyperfine coupling of approximately 400~kHz; see Extended Data Fig.~1.

\begin{figure}[htb]
	\centering
	\includegraphics[width=80mm]{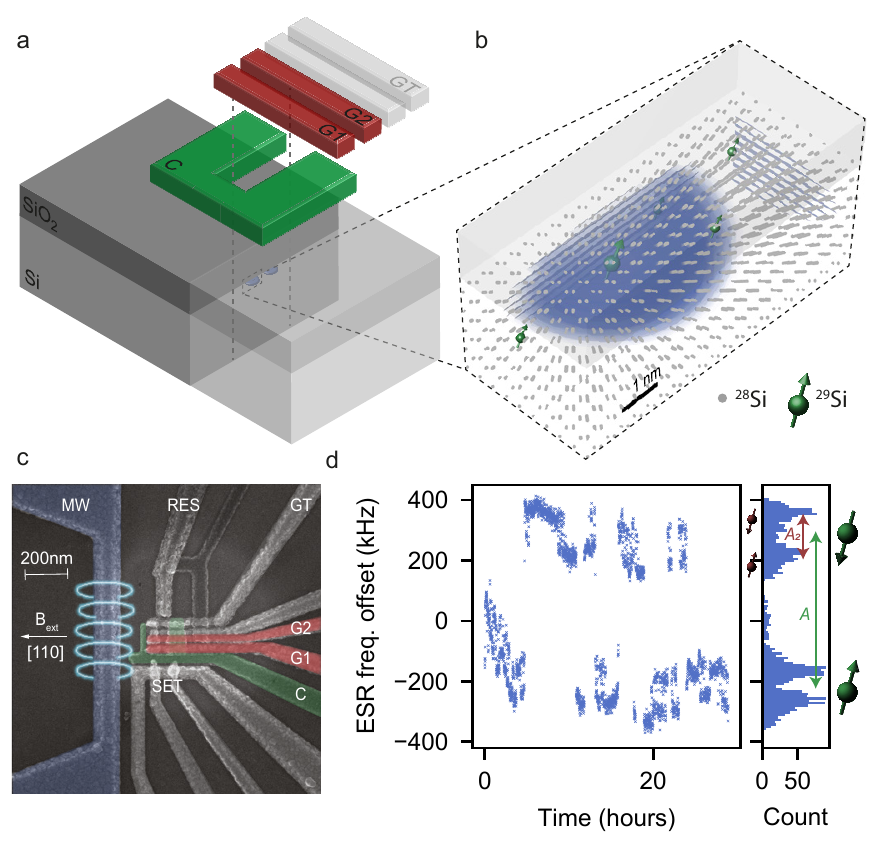}
	\caption{\label{fig:fig1} \textbf{Resolvable hyperfine coupling in a silicon metal-oxide quantum dot | a,} Schematic device layout, consisting of a double quantum dot formed under accumulation gates G1,G2, and laterally confined by gate C. GT controls the tunnel coupling to a nearby electron reservoir. \textbf{b,} Close-up of the interface region, showing an electron wavefunction (7~nm diameter) with vertical valley oscillations (blue lobes), projected over the silicon lattice (grey dots). The overlap of the wavefunction with nearby $\TNSi$ spins (green), shown here with random locations at 800~ppm density, determines the respective hyperfine coupling rates\cite{Assali2011Hyperfine}. Simulations yield the probability to find a nuclear spin with a particular hyperfine coupling rate (see Extended Data Fig.~1). Shown is one instance where the larger green spin indicates a 450~kHz coupled location. \textbf{c,} Scanning electron micrograph of the device layout, additionally showing the reservoir accumulation gate (RES), nearby single electron transistor (SET) to determine electron occupation and on-chip antenna (MW) to drive electron or nuclear spin resonance. Details of the device layout are described in Huang \emph{et al}\cite{Huang2019Fidelity}. \textbf{d,} (left) When monitoring the electron spin resonance (ESR) centre frequency, extracted by fitting repeated ESR frequency scans, bi-modal jumps can be observed on a timescale of order hours. (right) A histogram of the centre frequencies reveals the presence of a coupled nucleus (green spin), with hyperfine coupling $|A| \approx 450$~kHz, and a second $|A_2| \approx 120$~kHz coupled nucleus (red spin). Histogram bin-width is 8~kHz.}
\end{figure}

In this work we experimentally investigate the effect of individual nuclear spins on the operation of a double quantum dot device (Fig.~1c), that was previously characterised in Ref.\cite{Huang2019Fidelity}. The quantum dots QD1 and QD2 can be completely emptied, and single electrons can be loaded from the nearby electron reservoir. Using an external magnetic field $B_\mathrm{ext} = 1.42 \pm 0.04$~T to split the electron spin eigenstates by 39~GHz allows spin readout via the spin-selective unloading\cite{Elzerman2004Single-shot} of an electron from QD2. Furthermore, a single electron can be transferred between QD1 and QD2, while maintaining its spin polarisation\cite{Veldhorst2015two-qubit,Huang2019Fidelity}. Electron spin resonance (ESR) pulses applied to an on-chip microwave antenna allow coherent manipulations of the electron spin, with intrinsic spin transition linewidths around 50~kHz\cite{Veldhorst2014addressable}. When monitoring the spin resonance frequency ($f_\mathrm{e}$) for an electron loaded in QD1 (with QD2 empty) over extended periods of time, discrete jumps can be observed (Fig.~1d, left). Indeed, the histogram in Fig.~1d (right) suggests the presence of two distinct two-level systems, resulting in shifts in $f_\mathrm{e}$ of approximately 120~kHz and 450~kHz.

\begin{figure*}[p]
	\centering
	\includegraphics[width=80mm]{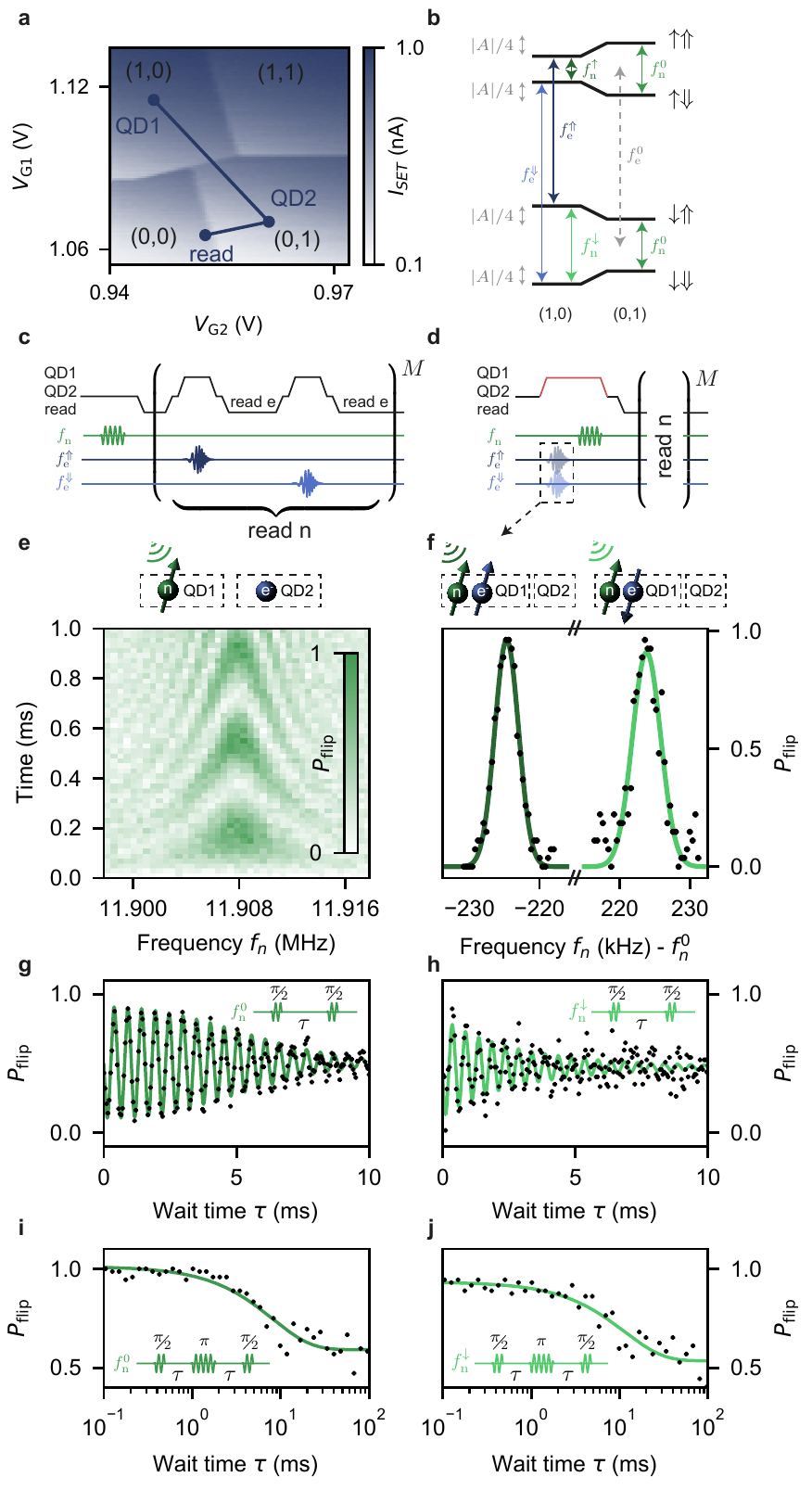}
	\caption{\label{fig:fig2} \textbf{$\TNSi$ nuclear spin qubit control and readout | a,} Double quantum dot charge stability diagram as a function of voltages $V_\mathrm{G1}$,$V_\mathrm{G2}$ applied to gates G1,G2. Shown are the electron occupation numbers (N1,N2), and voltage operation points used throughout for operating with a single electron in dot QD1 or QD2 as well as for the electron spin read-out (read). \textbf{b,} Energy levels of the joint electron-nuclear spin system, for a single nuclear spin coupled to quantum dot QD1. When QD1 is empty (right), the transition frequencies correspond to the bare Larmor frequencies $f_\mathrm{e}^0 = \left|\gamma_e B_\mathrm{ext}\right|$ and $f_\mathrm{n}^0 = \left|\gamma_\mathrm{Si} B_\mathrm{ext}\right|$. When the electron is loaded onto QD1 (left), the hyperfine interaction causes each frequency to split, depending on the state of the other spin. \textbf{c,} In order to detect the state of the nuclear spin, we compare the electron inversion probability around $f_\mathrm{e}^\nup$ and $f_\mathrm{e}^\ndo$. We can apply NMR pulses with QD1 unloaded, as in diagram \textbf{c,} or, as shown in diagram \textbf{d,} with QD1 loaded with a spin-$\edo$ electron. By applying an adiabatic ESR inversion (pulses shown in dotted square) we can also load QD1 with a spin-$\eup$ electron. \textbf{e,} In the unloaded case, schematically shown on (top), mapping the probability $P_\mathrm{flip}$ that a switch between $f_\mathrm{e}^\nup$ and $f_\mathrm{e}^\ndo$ has occurred as a function of applied NMR frequency and duration, we find (bottom), coherent oscillations of the nuclear spin. \textbf{f,} Loading a spin-$\edo$ electron, schematically shown on (top right), we find (right), the nuclear resonance frequency has shifted by $+|A|/2$. If we first flip the electron spin to $\eup$ (top left), we find (left), the nuclear frequency at $-|A|/2$, as expected. \textbf{g-j,} We perform nuclear Ramsey (\textbf{g,h}) and Hahn echo (\textbf{i,j}) sequences with the electron unloaded (\textbf{g,i}) and loaded on QD1 (\textbf{h,j}), in order to characterise the nuclear coherence properties. We find $T_2^\mathrm{*, unloaded} = 6.5 \pm 0.3$~ms, $T_2^\mathrm{Hahn, unloaded} = 16 \pm 2$~ms, $T_2^\mathrm{*, loaded} = 2.9 \pm 0.7$~ms and $T_2^\mathrm{Hahn, loaded} = 23 \pm 4$~ms. Ramsey sequences are performed detuned in order to accurately determine the decay. Above values correspond to a 1~hour integration time, see Extended Data Table~I for details.}
\end{figure*}

In order to determine whether the two-level fluctuations can be attributed to $\TNSi$ nuclear spins, we focus our attention on the 450~kHz shift. We first apply a radiofrequency (RF) tone with quantum dot QD1 unloaded, and then check whether the ESR frequency has shifted, by repeatedly probing the electron spin inversion probability around $f_\mathrm{e}^\nup = f_\mathrm{e}^0 - 225$~kHz or $f_\mathrm{e}^\ndo = f_\mathrm{e}^0 + 225$~kHz, with $f_\mathrm{e}^0$ the average ESR frequency. In Fig.~2e we show the probability $P_\mathrm{flip}$ that the electron spin resonance frequency has switched between $f_\mathrm{e}^\nup$ and $f_\mathrm{e}^\ndo$ after applying an RF pulse with varying frequency and duration, and find coherent oscillations centred around $f_\mathrm{n}^0 = 11.9078$~MHz. This frequency corresponds to a gyromagnetic ratio with magnitude $8.37 \pm 0.15$~MHz/T (where the uncertainty comes from the accuracy with which we can determine the applied field $B_\mathrm{ext}$), consistent with the bulk $\TNSi$ gyromagnetic ratio\cite{Pla2014Coherent} of $\gamma_\mathrm{Si} = -8.458$~MHz/T. We therefore conclude that the electron in quantum dot QD1 couples to a $\TNSi$ nucleus with hyperfine coupling $|A| = |f_\mathrm{e}^\nup- f_\mathrm{e}^\ndo| \approx 450$~kHz. We describe the joint spin system by a Hamiltonian of the form 
\begin{equation}
	H = -B_\mathrm{ext} (\gamma_e S_z + \gamma_\mathrm{Si} I_z) + A(S \cdot I),
\end{equation}
where $S$, $I$ are the electron and $\TNSi$ spin operators and $\gamma_e = -28$~GHz/T, the electron gyromagnetic ratio. Here, the interaction is dominated by the contact hyperfine term, with dipole-dipole terms expected to be several orders of magnitude smaller\cite{Assali2011Hyperfine}. The Hamiltonian results in energy eigenstates shown in Fig.~2b, with both electron and nuclear spin transitions splitting by A when the electron is loaded onto QD1. Repeating the same experiment as above, but applying the nuclear magnetic resonance (NMR) pulses after loading QD1 with a spin-down electron, we find the NMR frequency has shifted to $f_\mathrm{n}^\edo = f_\mathrm{n}^0 + |A|/2$, see Fig.~2f (right peak), yielding an accurate measurement of $A = -448.5 \pm 0.1$~kHz at this control point. Finally, we repeat the experiment once more, where we load QD1 with an electron with spin up (by applying adiabatic ESR inversion, see Methods), and confirm $f_\mathrm{n}^\eup = f_\mathrm{n}^0 - |A|/2$, see Fig.~2f (left peak). In Extended Data Fig.~3 we present results for another $\TNSi$ nuclear spin coupled to the electron spin in quantum dot QD2 that has hyperfine $A_\mathrm{QD2} = -179.8 \pm 0.2$~kHz. Electrostatic modelling based on the device gate geometry indicates an electron wavefunction diameter of around 8~nm, which is consistent with the hyperfine couplings observed in QD2, however the coupling observed for the $\TNSi$ in QD1, while possible for a 7~nm diameter, suggests that an additional disorder potential may have reduced the size of this dot wavefunction.

Having confirmed the ability to controllably address individual nuclear spins, we proceed to characterise a qubit encoded in this new resource. As observed from the interval between jumps in Fig.~1b, the nuclear spin lifetime (while repeatedly probing the electron resonance frequency, a process which likely affects that flip rate) extends to tens of minutes. By fitting an exponential decay to the intervals between the 450~kHz and 120~kHz jumps for the data in Fig.~1b, we find  $T_1^\mathrm{(450 kHz)} = 1.0 \pm 0.5$~hours and $T_1^\mathrm{(120 kHz)} = 10.0 \pm 0.6$~minutes. (see Supplementary Fig.~S1 for details). This means we can perform multiple quantum-non-demolition measurements of the nuclear spin state to boost the nuclear spin state readout fidelity\cite{Pla2013High-fidelity,Neumann2010Single-Shot}. A simple simulation (Methods) for $M$ repeated readouts, taking into account the 8~ms measurement cycle and the electron spin readout visibility of 76\% results in an optimal number of readouts $M_\mathrm{opt} = 26$, and an obtainable nuclear spin readout fidelity of 99.99\%. In this work we limit $M$ to 20, resulting in a measured nuclear spin readout fidelity of 99.8\% for the dataset in Fig.~3, see Methods for details. We determine the nuclear spin coherence times by performing nuclear Ramsey and Hahn-echo sequences, with the electron unloaded (Fig.~2g,i) and loaded (see Fig.~2h,j). We find nuclear coherence times between two and three orders of magnitude longer than those measured for the electron in this device ($T_2^{*\mathrm{,e}} \approx 15 \mathrm{\mu s}$\cite{Huang2019Fidelity}), but shorter than previously measured for nuclear spins coupled to donors in enriched silicon\cite{Muhonen2014Storing}, possibly due to the closer proximity of the device surface. A full overview of the measured coherence times is presented in Extended Data Table~I. In Supplementary Note~1, we discuss possible sources of dephasing resulting in the measured coherence times.

\begin{figure}[h!bt]
	\centering
	\includegraphics[width=80mm]{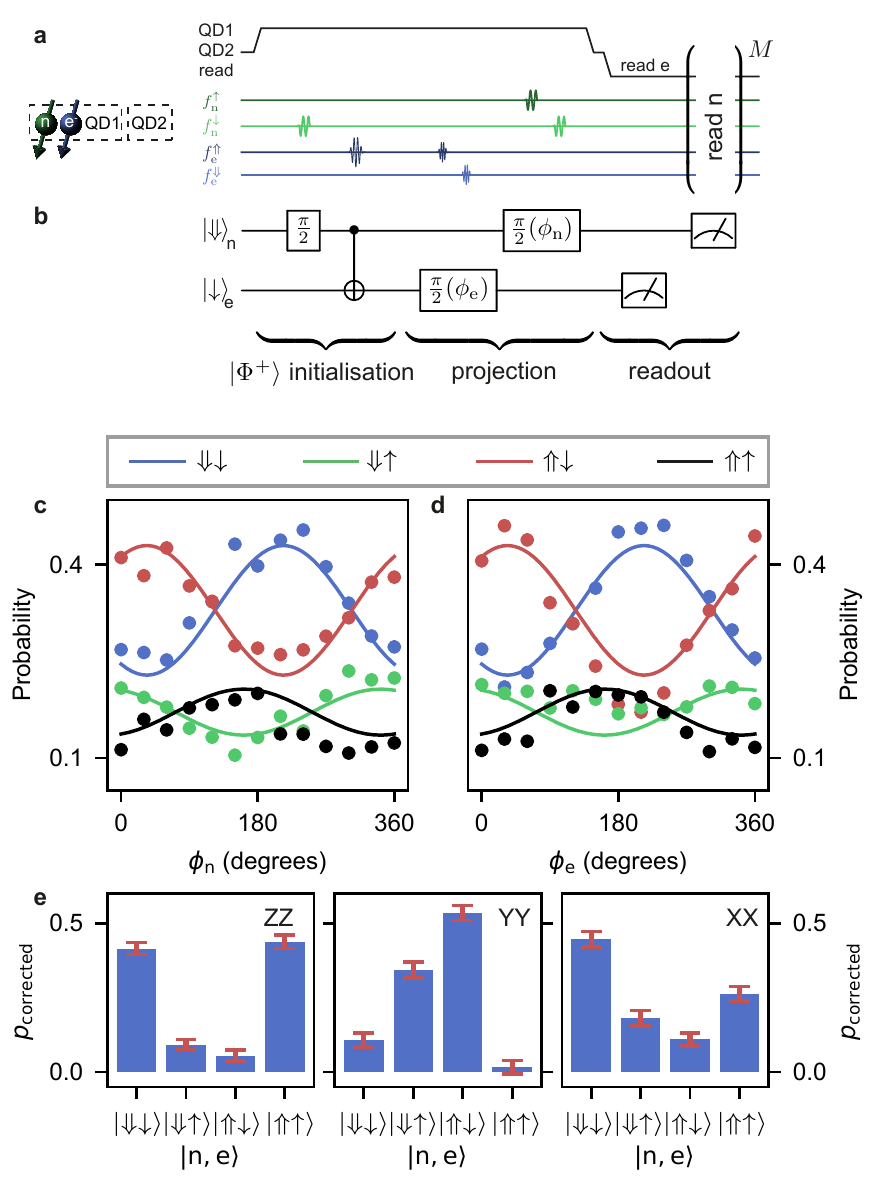}
	\caption{\label{fig:fig3} \textbf{Demonstration of entanglement between nuclear and electron spin state | a,} Pulse sequence used to implement the quantum circuit in \textbf{b,} to prepare the maximally entangled Bell state $\left|\Phi^+\right\rangle$, followed by projection onto the X-Y plane of both qubits and readout of the joint electron-nuclear spin state. All pulses used are coherent $\pi$ (controlled-NOT) or $\pi/2$ rotations (others). \textbf{c,d,} Varying the nuclear and electron projection phases $\phi_\mathrm{n}$, $\phi_\mathrm{e}$ respectively, we observe oscillations of the two-qubit parity, as expected for the initialised Bell state $\left|\Phi^+\right\rangle$. There is a clear difference in observed amplitude and phase for the electron spin-$\eup$ readout probabilities. The phase offset is caused by an AC stark shift induced by the off-resonant conditional ESR pulses used in the final state projection, while the reduced amplitude is a consequence of the asymmetric electron spin readout fidelity for electron spin-$\eup$ ($\approx 65\%$) versus spin-$\edo$ ($\approx 95\%$). Solid lines show the result of two sinusoidal fits, one for electron spin-$\eup$ (red, black) and one for electron spin-$\edo$ (green,blue). The data in \textbf{c,d,} and Extended Data Fig.~3a,b, are jointly fit, by including $180^\circ$ phase offsets depending on the initial and final nuclear spin states. \textbf{e,} We characterise the Bell state initialisation fidelity by measuring the joint state in bases XX, YY, ZZ. For the ZZ basis, the two-qubit state is measured without applying the projection pulses in \textbf{b,} while for the XX, YY bases we use the phases calibrated from \textbf{c,}. Shown are the joint readout probabilities, corrected for final electron readout fidelity.}
\end{figure}

In Fig.~3, we use coherent control to prepare entangled states of the joint electron-nuclear two-qubit system. We perform all operations with the electron loaded and construct the required unconditional rotations from two consecutive conditional rotations\cite{Dehollain2016Bells}, see Fig.~3a,b. After careful calibration of the AC Stark shifts induced by the off-resonant conditional ESR pulses (Fig.~3c,d), we characterise the prepared Bell state fidelity by measuring the two-qubit expectation values $\langle\mathrm{XX}\rangle$, $\langle\mathrm{YY}\rangle$ and $\langle\mathrm{ZZ}\rangle$ of the joint x,y,z-Pauli operator on the nucleus and electron, respectively. We correct the two-qubit readout probabilities for final electron readout errors only, which we calibrate by interleaving the entanglement measurement with readout fidelity characterisations (Methods). Note that because the nuclear spin is initialised by measurement here, we obtain an identical dataset corresponding to initial nuclear state $\left|\nup\right\rangle$, which we show in Extended Data Fig.~3. We find an average Bell state preparation fidelity of $73.0 \pm 1.9\%$. 

Our entanglement protocol is affected by several errors. By simulating our protocol, we estimate that the dominant source of error is the electron $T_2^{*\mathrm{,e}}$ ($\approx 10\%$), followed by the uncontrolled 120~kHz coupled $\TNSi$ nucleus, observed in Fig.~1b (5\%). Depending on the state of this nuclear spin, detuned ESR pulses cause an unknown phase shift. Other contributions include pulse duration calibration errors ($\approx2\%$) and the reduced NMR control fidelity with QD1 loaded (3\%).

A unique feature of our quantum-dot-coupled nuclear spin qubit is the large ratio of interdot tunnel coupling tc to hyperfine coupling A, so that $|t_c| \gg |A| \gg 1/T_2^{*\mathrm{,e}}$, with $|t_c|$ on the order of 1~GHz in this device\cite{Huang2019Fidelity}. We should therefore be able to accurately and adiabatically control the movement of an electron charge between neighbouring dots, inducing a predicable phase shift to the nuclear spin state. Importantly, the $\TNSi$ atom is isoelectronic with the rest of the crystal: its presence does not introduce any additional electrostatic perturbations. This is in contrast with the case of donors, where the nuclear spin resource is accompanied by a sharp Coulomb binding potential, which must be accounted for when designing the charge shuttling protocols\cite{Wolfowicz201629,Pica2016Surface,Harvey-Collard2017Coherent}. We can experimentally verify that the adiabatic transfer of an electron preserves the nuclear spin coherence with the pulse sequence shown in Fig.~4a. The sequence comprises a nuclear Ramsey-based experiment, where the electron is loaded onto QD1 from QD2 for a varying amount of time during free precession, while keeping the total evolution time constant. As expected, the loading of the electron causes a phase accumulation set by the hyperfine strength, see Fig.~4b, but preserves the nuclear spin coherence. To quantify the phase error induced by a load-unload cycle, we perform repeated electron shuttling, see Fig.~4c-d. 

The nuclear phase preservation raises the prospect to entangle nuclei in separate quantum dots, mediated by the electron shuttling\cite{Skinner2003Hydrogenic}. For that to work, the electron spin state itself must also remain coherent during transfer\cite{Fujita2017Coherent}. By performing an electron Ramsey experiment, where the first $\pi/2$ pulse is driven with the electron in QD1 and the second $\pi/2$ pulse is driven with the electron in QD2, see Fig.~4e,f, we demonstrate that electrons do indeed preserve coherence, presently with modest transfer fidelities. 

The readout and control of nuclear spins coupled to quantum dots shown here presents us with a variety of future research possibilities. First, the nuclear spin qubits could form the basis for a large-scale quantum processor, where initialisation, readout and multi-qubit interactions are mediated by electron spins\cite{Skinner2003Hydrogenic}. Second, the nuclear spin qubits could be used as a quantum memory\cite{Freer2017single-atom} in an electron-spin-based quantum processor. Implementations of quantum error correcting codes may benefit from integrated, long-term quantum state storage.  In particular, lossy or slow long-range interactions between quantum dots, for example mediated by microwave photonic qubits\cite{Mi2018coherent,Samkharadze2018Strong}, could be admissible if supplemented by local nuclear spin resources for memory or purification\cite{Wolfowicz201629,Bennett1996Purification,Nickerson2013Topological}. Finally, the nuclear spin qubits can be used as a characterisation tool for electron spin-based qubits.  For example, in the present experiment, the confirmed existence of a nucleus with 450~kHz hyperfine coupling bounds the electron wavefunction diameter to under 8~nm, a conclusion difficult to draw with purely electrostatic calculations or electronic measurements.  Further characterisations of electron spin dynamics may be envisioned by mapping the electron spin state to the nuclear spin, and employing the nucleus as a high-fidelity readout tool\cite{Dehollain2016Bells}. 

Limitations include the extended control times for the nuclear spin, as well as the effects of long NMR pulses on electron spin readout (see Supplementary Note~2). Both limitations could be addressed by redesigning the RF delivery, for example using a global RF cavity. Presently, the $\TNSi$ are randomly distributed, resulting in a range of hyperfine couplings for each quantum dot. Although the probability of obtaining at least one addressable $\TNSi$ per quantum dot is large (Extended Data Fig.~1), implantation of $\TNSi$ nuclei in (possibly further enriched\cite{Mazzocchi201999992}) silicon host material would allow the design of an optimal interaction. This could be done via ion implantation\cite{Donkelaar2015Single}, requiring a relatively modest precision on the order of the size of the quantum dots, which is much more forgiving than the precision needed for direct donor-donor coupling.

In summary, we have demonstrated coherent control, entanglement and high-fidelity readout of a single $\TNSi$ nuclear spin qubit, embedded in a lithographically-defined silicon quantum dot. We find that inter-dot electron tunnelling preserves the electron and nuclear spin coherence. The combination of controllable nuclear spin qubits with the long-range interactions afforded by electrons in silicon quantum dots provides a powerful new resource for quantum processing.

\begin{figure*}[h!btp]
	\centering
	\includegraphics[width=160mm]{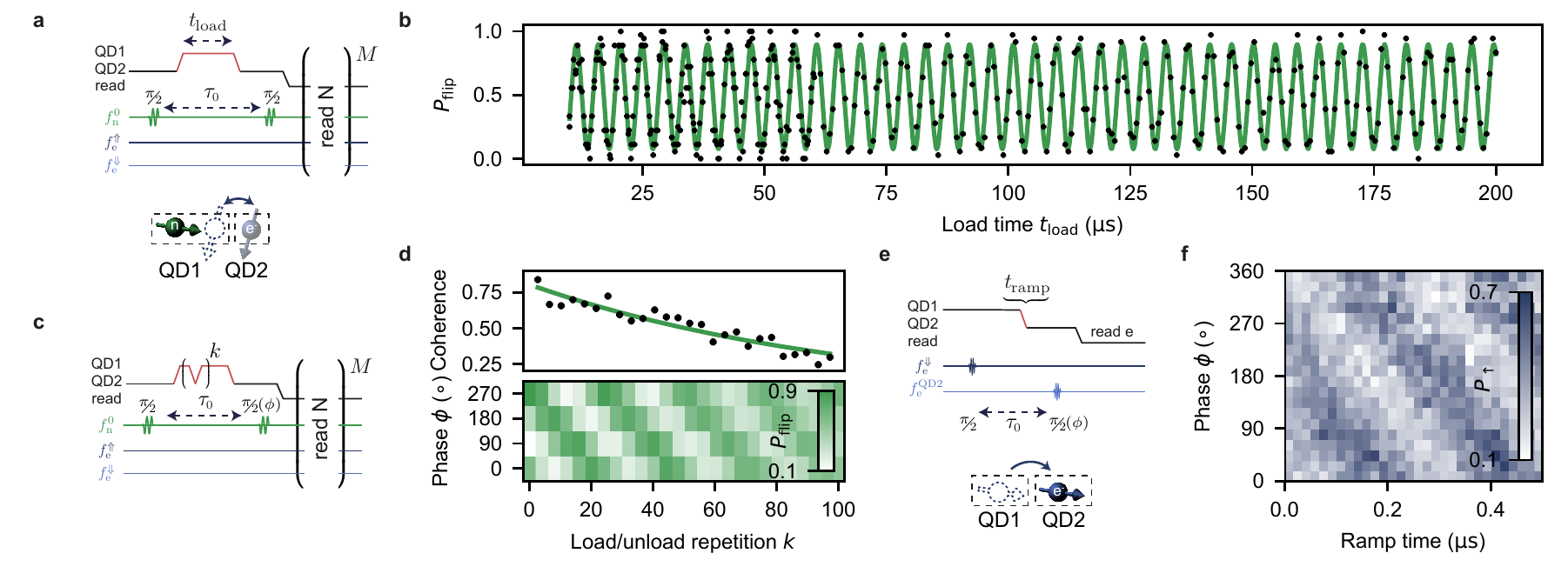}
	\caption{\label{fig:fig4} \textbf{Nuclear and electron spin coherence during electron transfer | a,} We verify that the nuclear spin coherence is maintained when moving the electron from quantum dot QD2 to QD1, by performing a non-detuned nuclear Ramsey experiment where we load the electron onto QD1 for a time $t_\mathrm{load}$ during free precession, while keeping the total precession time $\tau_0$ constant. \textbf{b,} With QD1 loaded the nuclear phase evolution is altered relative to the bare rotating frame by the hyperfine interaction, directly observable by the phase evolution as a function of $t_\mathrm{load}$ (keeping $\tau_0 = 0.5$~ms). The oscillation frequency yields another measurement for $|A|/2$. The oscillation visibility is limited by the electron spin initialisation fidelity (for electron spin-$\eup$ the oscillations have opposite phase). \textbf{c,} By repeatedly loading and unloading the electron we can estimate the loss of coherence due to the loading process. \textbf{d,} (bottom) To quantify the retained nuclear spin coherence independent of deterministic phase shifts we measure the probability $p_i$ of the nuclear spin state being in the states X, −X, Y, or –Y, corresponding to a spin up result after the final Ramsey pulse phases $\phi$ of $0^\circ$, $180^\circ$, $90^\circ$, $270^\circ$. (top) Nuclear coherence $C = \sqrt{(p_\mathrm{X}-p_{-\mathrm{X}})^2+(p_\mathrm{Y}-p_{-\mathrm{Y}})^2}$. Treating the loading/unloading process as a dephasing channel (Methods), we find an error probability per load/unload cycle of $0.45 \pm 0.29\%$. Here, $\tau_0 = 1.25$~ms is fixed. \textbf{e,} In an analogous measurement, we verify the electron spin coherence is maintained while shuttling it from QD1 to QD2, by performing an electron Ramsey where the first pulse is driven with the electron in QD1, and the second pulse with the electron in QD2. Note that because of the $g$-factor difference between quantum dots QD1 and QD2, the second pulse has a different frequency $f_\mathrm{e}^\mathrm{QD2}$. The nuclear spin state remains fixed in this experiment. \textbf{f,} Electron spin-$\eup$ readout probability as a function of final Ramsey phase $\phi$ and shuttling ramp time $t_\mathrm{ramp}$, showing a coherent spin transfer, with ~30\% visibility (no correction for readout). }
\end{figure*}

\bibliographystyle{naturemag}

\subsection{Acknowledgements}
We thank C.~Escott, P.~Harvey-Collard, V.V.~Dobrovitski for valuable discussions and feedback on the manuscript, and J.J.~Pla for technical assistance. We acknowledge support from the US Army Research Office (W911NF-17-1-0198), the Australian Research Council (CE170100012), Silicon Quantum Computing Proprietary Limited, and the NSW Node of the Australian National Fabrication Facility. The views and conclusions contained in this document are those of the authors and should not be interpreted as representing the official policies, either expressed or implied, of the Army Research Office or the U.S. Government. The U.S. Government is authorised to reproduce and distribute reprints for Government purposes notwithstanding any copyright notation herein. B.H. acknowledges support from the Netherlands Organisation for Scientific Research (NWO) through a Rubicon Grant. K.M.I. acknowledges support from a Grant-in-Aid for Scientific Research by MEXT.

\subsection{Author contributions}
B.H. and W.H. performed the experiments. K.W.C. and F.E.H. fabricated the devices. K.M.I. prepared and supplied the 28Si wafer. B.H., W.H., C.H.Y., J.Y., T.T.  and A.L. designed the experiments. B.H., W.H. and J.Y. analysed the data. T.D.L. performed hyperfine and coherence simulations. B.H. wrote the manuscript with input from all co-authors. A.M., A.L., and A.S.D. supervised the project.

\subsection{Methods}
\subsubsection{Experimental methods}
Details of sample fabrication and experimental setup can be found in Huang \emph{et al}\cite{Huang2019Fidelity}. NMR pulses were generated by a secondary microwave vector signal generator, combined with the ESR pulses using a resistive combiner.
Conditional ESR pulses for nuclear readout are performed via adiabatic inversions with a frequency span of $f_\mathrm{e}^\nup - 300$~kHz to $f_\mathrm{e}^\nup + 50$~kHz and $f_\mathrm{e}^\ndo - 50$~kHz to $f_\mathrm{e}^\ndo + 300$~kHz for reading nuclear spin $\nup$, $\ndo$ respectively. The adiabatic pulse has a duration of $650~\mathrm{\mu s}$ and a power corresponding to a 100~kHz Rabi frequency.
Electron spin transfers between QD1 and QD2 are performed with a $1 \mathrm{\mu s}$ linear ramp, except where indicated otherwise.

\subsubsection{Nuclear spin readout fidelity}
We model the repetitive nuclear readout as a stochastic process, where, as a function of the number of shots $M$, the fidelity is limited by the nuclear $T_1$ decay on the one hand,
\begin{equation}
F_{T_1}= \exp\left[-2M t_\mathrm{shot}/T_1\right],
\end{equation}
with $t_\mathrm{shot} = 8$~ms the measurement time per shot, $T_1 = 1$~hour, and the factor 2 comes from the fact that for each shot we read out the electron twice; once for an inversion around $f_\mathrm{e}^\nup$ and once for $f_\mathrm{e}^\ndo$.  On the other hand the fidelity is limited by the cumulative binomial distribution representing the majority voting of M single shots:
\begin{equation}
	F_\mathrm{shot}= \sum_{k=0}^{2M/2} \binom{2M}{k}  (1-F_\mathrm{e}^\mathrm{avg})^k (F_\mathrm{e}^\mathrm{avg})^{2M-k}.
\end{equation}	
A first order estimate for the nuclear readout is then obtained by 
\begin{equation}
	F_\mathrm{n}\approx (1-F_{T_1})F_\mathrm{shot}+ F_{T_1} (1-F_\mathrm{shot}),
\end{equation}
resulting in a minimum infidelity $1-F_\mathrm{n} = 10^{-4}$ for $M_\mathrm{opt} = 26$, for $F_\mathrm{e}^\mathrm{avg} = 76.5\%$, where we have taken the average electron spin readout fidelity recorded for the dataset in Fig.~3e.

\subsubsection{Nuclear-electron entanglement experiment: experimental details}
For the datasets presented in Fig.~3e and Extended Data Fig.~3c, we interleave the following measurement sequences:
\begin{enumerate}
	\item Bell state preparation + ZZ projection
	\item Bell state preparation + XX projection
	\item Bell state preparation + YY projection
	\item Electron spin-$\edo$ readout characterisation for ZZ projection
	\item Electron spin-$\eup$ readout characterisation for ZZ projection
	\item Electron spin-$\edo$ readout characterisation for XX and YY projection
	\item Electron spin-$\eup$ readout characterisation for XX and YY projection
\end{enumerate}
Prior to running each sequence, we initialise the electron spin state to $\edo$, using a spin relaxation hotspot at the (0,1)-(1,0) charge transition (details in \emph{et al.}\cite{Huang2019Fidelity}). After running a sequence once, we read the state of both electron and nuclear spin (corresponding to a total of $2M+1$ electron spin readouts). We then perform an ESR frequency check and, if necessary, calibration. The frequency check proceeds by applying a weak, resonant ESR pulse (60~kHz Rabi frequency), and fails if the spin inversion probability drops below 0.3. If the check fails for both $f_\mathrm{e}^\nup$ and $f_\mathrm{e}^\ndo$, the ESR frequency is recalibrated using a series of Ramsey sequences to estimate the detuning. Details of this ESR frequency calibration are described in the Supplementary Information of Huang \emph{et al}\cite{Huang2019Fidelity}. After recording 10 datapoints of sequence 1 in this manner, we switch to sequence 2, and so forth. After sequence 7 we loop back to sequence 1, until the end of the measurement. The nuclear spin initialisation is given by the readout result in the previous sequence. Total measurement time for the presented dataset was 9.5~hours, resulting in 4320 Bell state preparations.

The aim of sequence 4-7 is to record the actual average electron spin readout fidelity while recording the dataset. To estimate the spin-$\edo$ readout fidelity (sequence 4,6), we apply no ESR pulses and measure the spin-$\edo$ readout probability. To estimate the spin-$\eup$ readout fidelity we apply an adiabatic inversion of the electron spin, consisting of a $650~\mathrm{\mu s}$ long 2.8~MHz wide frequency sweep centred around $f_e^0$, with a power corresponding to a 100~kHz Rabi frequency, and measure the spin-$\eup$ readout probability. Sequence 4,5 each have the same NMR pulses applied as sequence 1, but applied far detuned, in order to mimic the effect of the ZZ-projection NMR pulses on the electron spin readout fidelity, while not changing the nuclear spin state itself. Similarly, sequence 6,7 have the same NMR pulses as sequence 2, but far detuned, to mimic the effect of XX, YY-projection pulses. For the ZZ-projection we find readout fidelities $F_e^\edo = 88.4\%$ and $F_e^\eup = 73.3\%$, while for the XX, YY-projection we find $F_e^\edo = 80.7\%$ and $F_e^\eup = 67.4\%$. See also Supplementary Note~2 for a discussion of the effect of RF pulses on electron spin readout fidelity.

Finally, data from sequence 4-7 is also used to obtain an estimate for the nuclear spin readout fidelity: since all NMR pulses are applied off-resonant, the nuclear spin should remain unchanged. If the nuclear spin is read out differently after running sequence 4-7, this indicates a readout error has occurred. We find 5 readout errors in 3200 nuclear spin readouts, identified as such by a single outcome being different in a sequence of 10. The readout fidelity is estimated as the fraction of readout errors.

Bar plots shown in Fig.~3e and Extended Data Fig.~3c are corrected using their respective electron spin readout fidelity characterisation, using direct inversion. We estimate the Bell state fidelity using $F = F_{\langle ZZ \rangle} /2 + F_{\langle YY \rangle}/2 + F_{\langle XX \rangle} /2 - 1/2$, where $F_{\langle ZZ \rangle} = p_{\ndo\edo}+p_{\nup\eup}$, $F_{\langle YY \rangle} = p_{\ndo\eup}+p_{\nup\edo}$, $F_{\langle XX \rangle} = p_{\ndo\edo}+p_{\nup\eup}$ for nuclear spin-$\ndo$ initialised data (Fig.~3e), and $F_{\langle YY \rangle} = p_{\ndo\edo}+p_{\nup\eup}$, $F_{\langle XX \rangle} = p_{\ndo\eup}+p_{\nup\edo}$ for nuclear spin-$\nup$ initialised data (Extended Data Fig.~3c). 

\subsubsection{Nuclear-electron entanglement: error analysis}
Using the two-spin Hamiltonian, eq. (1), with two control fields $V_\mathrm{ESR}=|\gamma_\mathrm{e}| B_1(t)S_x$  and $V_\mathrm{NMR}=|\gamma_\mathrm{Si}| B_1(t) I_x$, and taking the secular and rotating wave approximation, we perform a time evolution simulation to estimate the effects of various noise sources on the nuclear-electron Bell state fidelity. We simulate the exact control sequences $B_1(t)$ used in the experiment. The simulation calculates the operator at any specific time $U_{dt}(t) = \exp\left[-2i\pi H_\mathrm{RWA}
(t) t\right]$ resulting in a final operator $U= \prod U_{dt}(t)$. We incorporate quasi-static noise along $I_x$, $I_z$ and $S_z$ directions following a Gaussian distribution with standard deviation of $\frac{1}{\sqrt{2}\pi T_2^\mathrm{Rabi,n}}$, $\frac{1}{\sqrt{2}\pi T_2^\mathrm{*,n}}$,  $\frac{1}{\sqrt{2}\pi T_2^\mathrm{*,e}}$ respectively, and repeat the simulation for 1000 times to obtain the average final measurement probabilities. We use values $T_2^\mathrm{Rabi,n} = 1.1$~ms,  $T_2^\mathrm{*,n} = T_2^\mathrm{*,loaded} = 2.9$~ms, $T_2^\mathrm{*,e} = 15~\mathrm{\mu s}$. The value for $T_2^\mathrm{*,e}$ has a large uncertainty, ranging from 8 to $22 \mathrm{\mu s}$ depending on the exact ESR frequency feedback settings and interval\cite{Huang2019Fidelity}. To simulate the effect of the uncontrolled 120~kHz coupled $\TNSi$ spin, we estimate the probability that the nuclear spin flips within the time between ESR frequency checks, resulting in an unnoticed frequency shift. Using $T_1^\mathrm{(120 kHz)} = 10$~min, and an average time between ESR frequency checks of 40~seconds, we find a probability of 7\% of running the entanglement sequence with 120~kHz detuned ESR pulses. Finally, to simulate the effect of pulse calibration errors, we estimate our pulse-length calibration is accurate within 5\%. Error percentages quoted in the Main text are the reduction in final Bell state fidelity resulting from incorporating the corresponding error mechanism only, with all other error mechanisms turned off in the simulation.

\subsubsection{Coherent loading dephasing analysis}
We model the effect of transferring the electron between QD1 and QD2 on the nuclear spin state as a dephasing channel
\begin{equation}
\rho=\begin{pmatrix}
\rho_{00} & \rho_{01} \\
\rho_{10} & \rho_{11}
\end{pmatrix}
\rightarrow \rho' = (1-p_\mathrm{err})\rho + p_\mathrm{err}\begin{pmatrix}
\rho_{00} & 0 \\
0 & \rho_{11}
\end{pmatrix}.
\end{equation}
This model yields an exponentially decaying off-diagonal matrix element magnitude as a function of channel transfers $k$, $|\rho_{01}(k)| = 1/2 \exp{-k*p_\mathrm{err}} = 1/2 C(k)$, which is the measured coherence defined in the caption of Fig.~4d,e.

\clearpage
\onecolumngrid
\appendix
\subsection{Extended Data}

\renewcommand\figurename{Extended Data Figure}   
\renewcommand\thefigure{\thesection \arabic{figure}}   
\setcounter{figure}{0} 
\renewcommand\thetable{\thesection \Roman{table}}  
\renewcommand\tablename{Extended Data Table}   

\begin{figure}[h!]
	\centering
	\includegraphics[width=80mm]{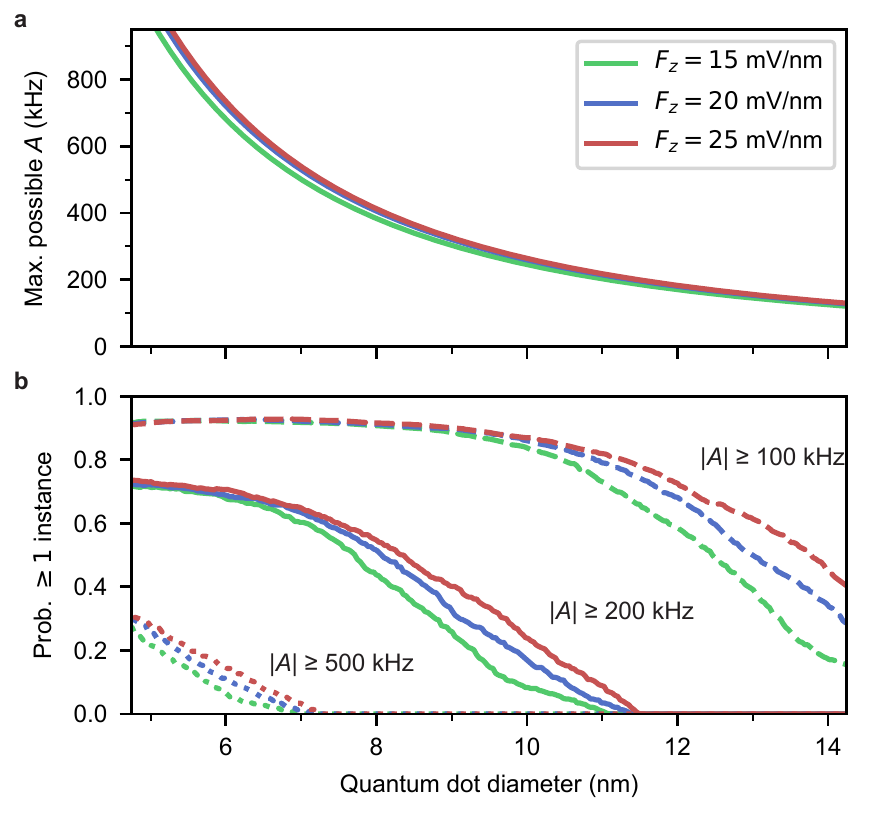}
	\caption{\label{fig:extfig1} \textbf{Expected hyperfine interaction | a,} Maximum possible hyperfine interaction for given dot diameter, defined as the $1/\mathrm{e}$ point of the envelope charge distribution, and vertical confining electric field $F_z$. \textbf{b,} Probability of observing at least one nuclear spin with hyperfine coupling $|A| \geq 100$~kHz (dashed line), $|A| \geq 200$~kHz (solid line) and $|A| \geq 500$~kHz (dotted line), for 800~ppm $\TNSi$ material. These distributions are found by calculating the electron wavefunction density for an Airy envelope function\cite{Davies1997Physics}, sinusoidally oscillating in the vertical dimension due to valley oscillations, and with a transverse Gaussian shape, all superimposed over an unstrained silicon lattice. We assume $\TNSi$ are randomly placed with 800~ppm probability at each lattice site and evaluate the resulting hyperfine contact interaction\cite{Assali2011Hyperfine}. The probabilities grow with dot diameter due to increased number of sites overlapped, but then shrink for large dots due the reducing hyperfine contact at each site.}
\end{figure}

\begin{figure}[h!]
	\centering
	\includegraphics[width=80mm]{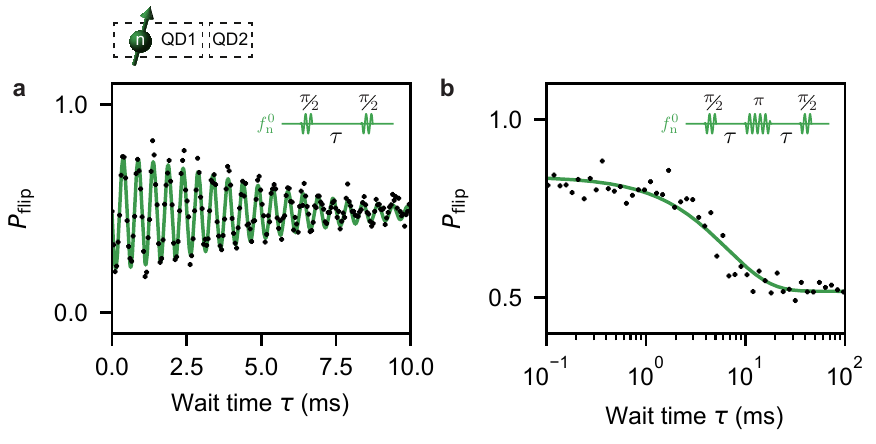}
	\caption{\label{fig:extfig2} \textbf{Ramsey and Hahn echo measurement in the (0,0) charge state | a,} Ramsey measurement and \textbf{b,} Hahn echo measurement with nuclear pulses and free evolution time in charge configuration (0,0). Resulting values $T_2^\mathrm{*,(0,0)} = 6.0 \pm 0.6$~ms and $T_2^\mathrm{Hahn,(0,0)} = 13.1 \pm 1.5$~ms are within error to those obtained for the unloaded-(0,1) charge configuration, see Fig.~2g,i.}
\end{figure}

\begin{figure}[h!]
	\centering
	\includegraphics[width=80mm]{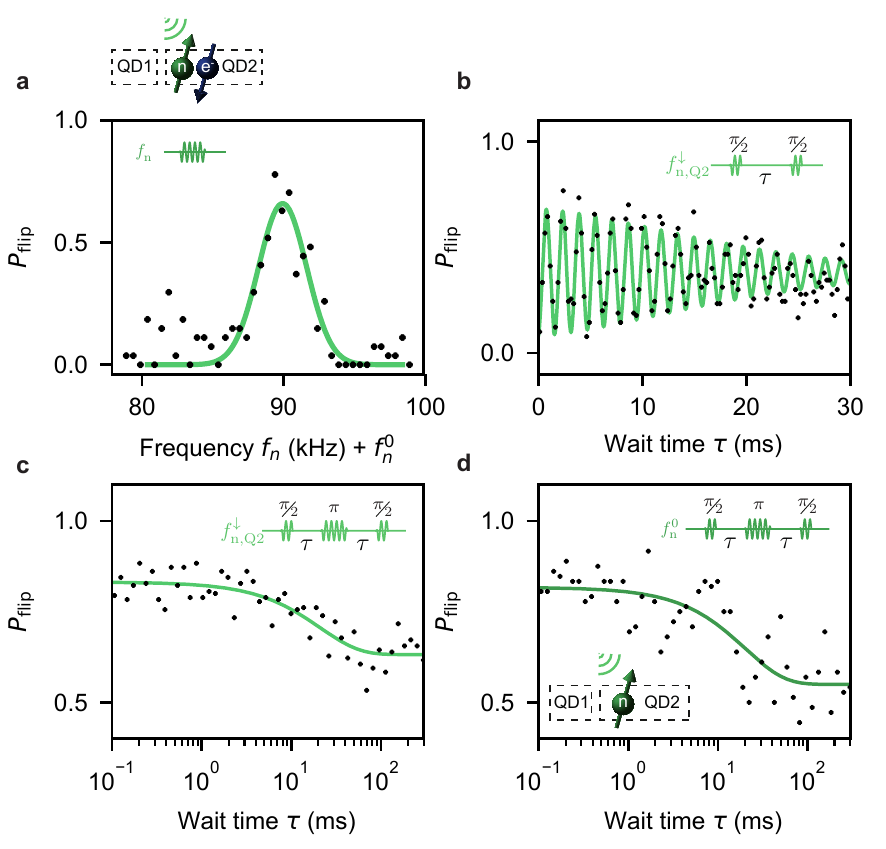}
	\caption{\label{fig:extfig3}  \textbf{Second $\TNSi$ qubit coupled to quantum dot QD2 | a,} NMR frequency scan with QD2 loaded with a spin $\edo$ electron, charge configuration (0,1), with a spin down electron reveals a $\TNSi$ nuclear spin coupled by $A_\mathrm{QD2} = -179.8 \pm 0.2$~kHz. Note that the nuclear spin readout contrast is reduced due to the small hyperfine splitting. \textbf{b,} Loaded Ramsey measurement yields $T_2^\mathrm{*,loaded} = 21 \pm 5$~ms, for 1 hour integration time. \textbf{c,} Loaded Hahn echo measurement yields $T_2^\mathrm{Hahn, loaded} = 42 \pm 11$~ms. \textbf{d,} unloaded, charge state (0,0), Hahn echo yields $T_2^\mathrm{Hahn,(0,0)} = 40 \pm 13$~ms.}
\end{figure}

\begin{figure}[h!]
	\centering
	\includegraphics[width=80mm]{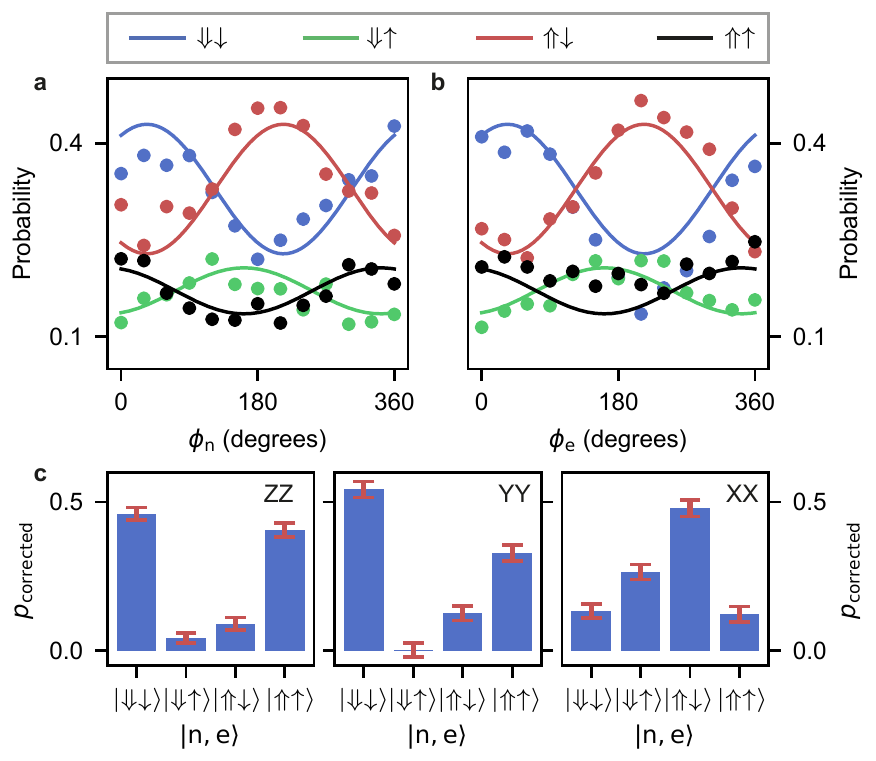}
	\caption{\label{fig:extfig4} \textbf{Nuclear-electron entanglement data for opposite nuclear spin initialisation | a,b,} As expected, for a nuclear spin-$\nup$-initialised state, varying the nuclear and electron projection phases $\phi_\mathrm{n}$, $\phi_\mathrm{e}$ respectively, we observe oscillations with opposite phase compared to those for a nuclear spin-$\ndo$-initialised state, compare Fig.~3c,d. \textbf{c,} Accordingly, XX and YY projections have opposite parity, compare Fig.~3e.}
\end{figure}

\begin{table*}[htb]
\begin{tabular}{@{}lllllllll@{}}
\toprule
\multirow{2}{*}{\begin{tabular}[c]{@{}l@{}}Nuclear spin\\ in quantum dot\end{tabular}}  & \multirow{2}{*}{Charge  state} & \multirow{2}{*}{\begin{tabular}[c]{@{}l@{}}Hyperfine\\ magnitude\end{tabular}} & \multicolumn{4}{l}{Ramsey}                                         & \multicolumn{2}{l}{Hahn echo}   \\
                                              &                                &                                       & $T_2^*$             & Exponent $\alpha$      & Int. time & Figure  & $T_2^\mathrm{Hahn}$   & Figure  \\ \colrule
QD1                                           & (0,1) -  unloaded              & 450 kHz                               & $6.6 \pm 0.2  $~ms  & $2.11 \pm 0.21$ 		  & 3.2 hrs   & Fig.~2g & $16.3 \pm 2.4  $~ms  	& Fig.~2i \\
QD1                                           & (1,0) -  loaded                & 450 kHz                               & $2.9 \pm 0.7  $~ms  & $0.91 \pm 0.25$ 		  & 1.1 hrs   & Fig.~2h & $22.8 \pm 4.1  $~ms  	& Fig.~2j \\
QD1                                           & (0,0) -  unloaded              & 450 kHz                               & $5.9 \pm 0.2  $~ms  & $1.54 \pm 0.39$ 		  & 7 hrs     & Ext.~2a & $13.1 \pm 1.5  $~ms  	& Ext.~2b \\
QD2                                           & (0,1) -  loaded                & 180 kHz                                & $21.3 \pm 2.0 $~ms  & $1.74 \pm 0.16$ 		  & 6.3 hrs   & Ext.~3b & $42.2 \pm 10.6 $~ms 	& Ext.~3c \\
QD2                                           & (0,0) -  unloaded              & 180 kHz                                & \multicolumn{4}{l}{Not  measured}                 					& $40.5 \pm 13.0 $~ms 	& Ext.~3d \\ \botrule

\end{tabular}
\caption{\label{tab:tab1} \textbf{Overview of nuclear coherence times |}  Details and fitted values for all Ramsey and Hahn echo sequences performed on two $\TNSi$ nuclear spins, one in quantum dot QD1, one in QD2, for different charge states. Ramsey values are fits to a sinusoidal function with envelope decay $~ \exp\left[-(\tau/ T_2^*)^\alpha\right]$. Int. time indicates total integration time for the measurement. Hahn echo values are fits to an exponential decay $~\exp\left[-2\tau/T_2^\mathrm{Hahn}\right]$. Figure panels displaying each measurement are indicated. Note that for the $T_2^*$-values given in the captions of Fig.~2 and Extended Data Figures~2,3, the integration times are all limited to 1~hour, for comparison.}
\end{table*}

\end{document}


\title{A silicon quantum-dot-coupled nuclear spin qubit - SUPPLEMENTARY INFORMATION}
\author{Bas~Hensen}
\thanks{These authors contributed equally}
\author{Wister~Huang}
\thanks{These authors contributed equally}
\author{Chih-Hwan~Yang}
\author{Kok~Wai~Chan}
\author{Jun~Yoneda1}
\author{Tuomo~Tanttu}
\author{Fay~E.~Hudson}
\author{Arne~Laucht}
\affiliation{Centre for Quantum Computation and Communication Technology,
School of Electrical Engineering and Telecommunications,
The University of New South Wales, Sydney, New South Wales 2052, Australia}
\author{Kohei~M.~Itoh}
\affiliation{School of Fundamental Science and Technology, Keio University, 3-14-1 Hiyoshi, Kohoku-ku, 
Yokohama 223-8522, Japan}
\author{Thaddeus~D.~Ladd}
\affiliation{HRL Laboratories, LLC, 3011 Malibu Canyon Rd., Malibu, CA, 90265, USA}
\author{Andrea~Morello}
\author{Andrew~S.~Dzurak}
\email{b.hensen@unsw.edu.au,wister.huang@unsw.edu.au, a.dzurak@unsw.edu.au}
\affiliation{Centre for Quantum Computation and Communication Technology,
School of Electrical Engineering and Telecommunications,
The University of New South Wales, Sydney, New South Wales 2052, Australia}

\maketitle

\section*{Supplementary Note 1: Discussion of measured dephasing times}
The 6.6 ms dephasing time observed for the strongly hyperfine-coupled $^{29}$Si nucleus in QD1 is shorter than expected from residual nuclear magnetisation. Even in isotopically natural silicon, dephasing due to neighbouring $^{29}$Si nuclei is of order 15 ms\cite{watanabe_29si_2003}; the 800 ppm silicon in the present sample should increase this timescale by a factor of at least the square root of the isotopic content to over 100~ms\cite{abe_electron_2010}. Previous measurements for $^{31}$P donor nuclei in 800ppm silicon[6] showed Hahn-echo times up to 1.8 seconds with the electron ionised.  One possible explanation for magnetic dephasing may be other types of nuclear spins; in particular the aluminium gates a small distance away have 100\% $^{27}\mathrm{Al}$ spins each with fluctuating spin-5/2 nuclear magnetisation.  In Supplementary Fig.~S2, we estimate the dephasing time due the dipolar magnetic field of these nuclear spins using the Van Vleck method of moments\cite{slichter_principles_1990}. We find that the influence of $^{27}\mathrm{Al}$ spins from the gate is comparable to the anticipated dephasing from 800ppm $^{29}$Si nuclei, and still too small to account for the observed value.  Other sources of local magnetic fields such as dangling-bond electrons at the oxide interface\cite{de_sousa_dangling-bond_2007} or Pauli ferromagnetism from the metal gates and baths are unlikely to contribute to $T_2^*$ and $T_2$ at the level observed at this magnetic field and temperature due to the high degree of electron spin polarisation anticipated, although nearly degenerate spin-pair flip-flops, either from nearby nearest-neighbour $^{29}$Si nuclei or trapped, localised electron spins in the oxide, may be responsible if randomly placed appropriately near the nucleus.  We note that other $^{29}$Si nuclei, such as that in QD2, exhibit longer $T_2^*$ and $T_2$ times as summarised in Extended Data Table I, with values close to that expected from fluctuating $^{29}$Si and $^{27}\mathrm{Al}$ nuclear magnetisation, suggesting a highly localised and random nuclear spin dephasing source in QD1.

\section*{Supplementary Note 2: RF induced degradation of electron spin readout fidelity} The application of RF pulses results in a reduced electron spin readout fidelity. This effect is mostly influenced by the average applied RF power and can largely be mitigated by decreasing the experimental duty cycle, while maintaining the same pulse power and duration, see Supplementary Fig.~S3. For higher applied RF powers, the probability to observe an electron tunnelling event for electron spin-$\uparrow$ is reduced. The effect is independent from the time-delay between applying an RF pulse and performing the electron spin readout. The effect also does not depend on the gate voltage operation point where the RF pulses are applied (in particular, the detuning of the dot chemical potential from any reservoir or inter-dot transitions). The physical mechanism of this effect is presently not well understood, however the strong dependence on experiment duty cycle suggests a macroscopic origin such as device chip heating, or local charge accumulation. For future experiments, the average RF power can be reduced by optimising the RF delivery, for example by employing an RF cavity.
\clearpage
\onecolumngrid
\section*{Supplementary Figures}

\begin{figure*}[hbt]
	\centering
	\includegraphics[width=160mm]{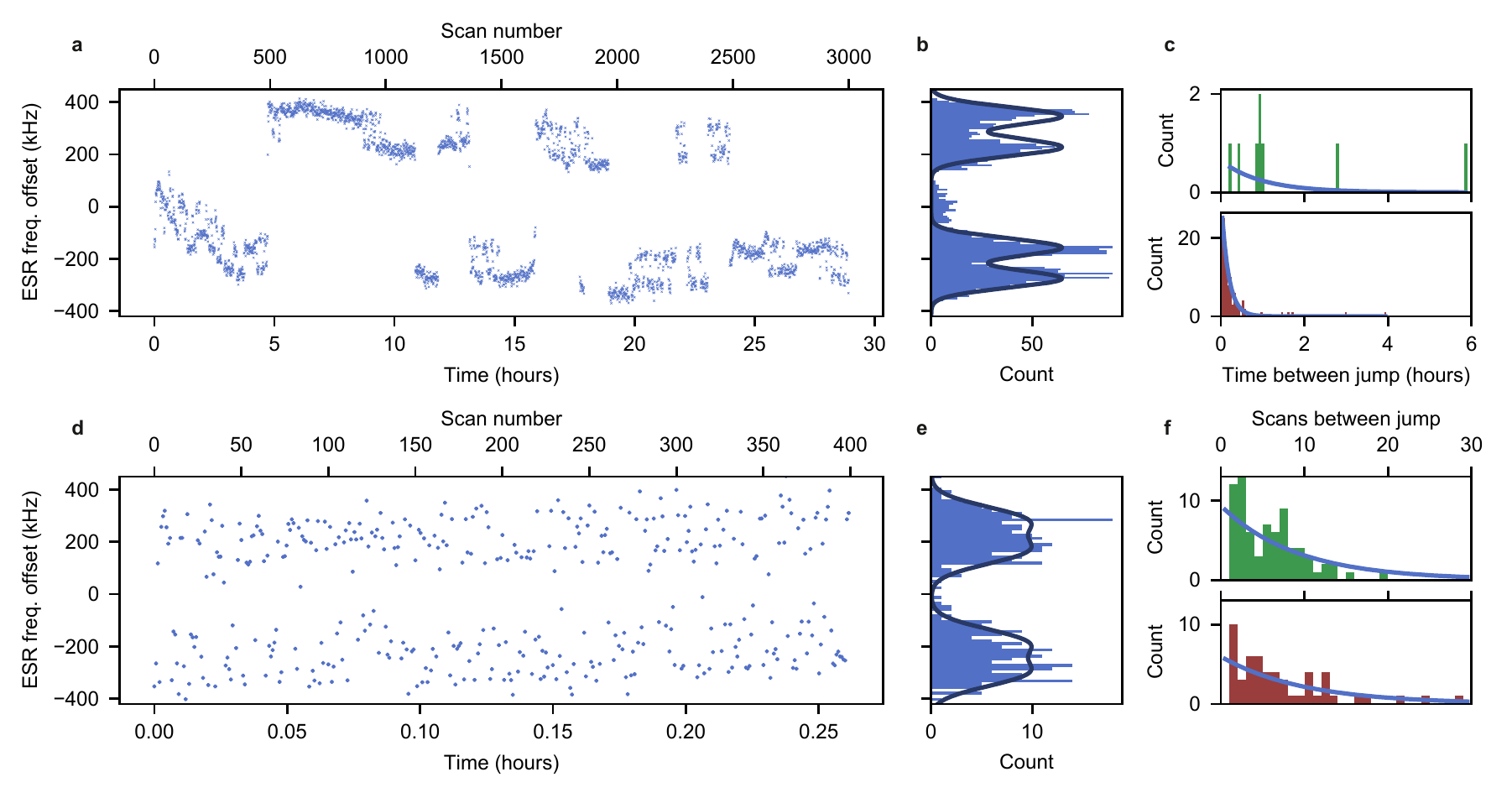}
	\caption{\label{fig:supfig1} \textbf{Natural and induced nuclear spin flips | a,}  ESR centre frequency, extracted by fitting repeated ESR frequency scans. This data is identical to the data shown in Fig. 1b of the Main text. \textbf{b,} Fitting the resulting histogram to the expected spectrum for two hyperfine coupled nuclei, a sum of four Gaussian peaks at frequencies, $f_0^\mathrm{e} \pm A_1 \pm A_2$, we obtain $|A_1| =  503 \pm 4$~kHz, $|A_2|=119 \pm 3$~kHz and a line-width $\sigma = 34 \pm 1.5$~kHz. \textbf{c,} We classify a frequency shift $\Delta f$ between two subsequent scans as a $A_1$-related shift when  $|A_1|-2\sigma \leq |\Delta f| \leq |A_1|+2\sigma$, and as a $A_2$ related shift for $|A_2|-\sigma \leq |\Delta f| \leq |A_2|+\sigma$. The histogram shows the occurrence of time-delays between $A_1$ (top) and $A_2$ (bottom) related frequency shifts. An exponential fit to the histogram give the values for $T_1^{A_1} = 1.0 \pm 0.5$~hours and $T_1^{A_2} = 10 \pm 0.6$ minutes. \textbf{d,e} We include a NMR-pulse resonant with the unloaded $^{29}$Si transition frequency $f_0^\mathrm{n}$ between each ESR frequency scan, with no electrons loaded during the pulse. A drastic increase in the rate of both $A_1$ and $A_2$ related frequency shifts can be observed. \textbf{f} Performing the same analysis as in \textbf{c,} (using the values for $|A_1|$, $|A_2|$, $\sigma$, from \textbf{b,}), yields an exponential decay of $9\pm2$ and $10\pm3$ scans for $A_1$, $A_2$ related shifts, corresponding to an induced $T_1$-like decay of about 20 seconds, demonstrating the $^{29}$Si spin origin of both $A_1$ and $A_2$ transitions.}
\end{figure*}

\begin{figure}[hbt]
	\centering
	\includegraphics[width=80mm]{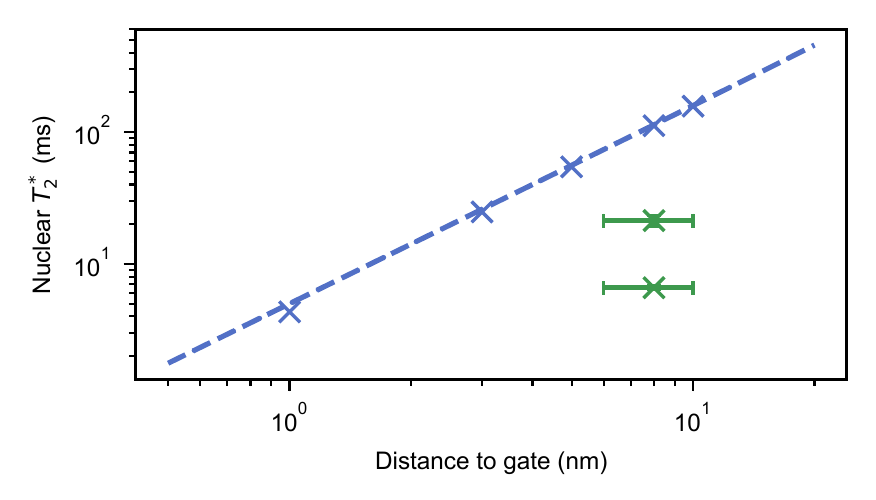}
	\caption{\label{fig:supfig2} \textbf{Expected dephasing time due to gate electrodes $^{27}$Al nuclear spins | } Using the Van~Vleck method of moments\cite{slichter_principles_1990}, we simulate the effect of a large aluminium gate some distance above the $^{29}$Si nucleus. The exact distance is uncertain due to the uncertain exact vertical placement of the $^{29}$Si nucleus and oxide thickness. We place the $^{27}$Al in the calculation according to FCC single-crystal aluminium metal for simplicity. Shown are the simulation results for a full summation for a 50 nm thick, 300 nm by 100 nm sized aluminium electrode area (blue crosses), as well as a approximate integral form, treating the gate as a cylindrical body (dashed blue line). Also shown are the measured values for QD1 and QD2 (green crosses), with estimated uncertainties.}
\end{figure}

\begin{figure}[hbt]
	\centering
	\includegraphics[width=80mm]{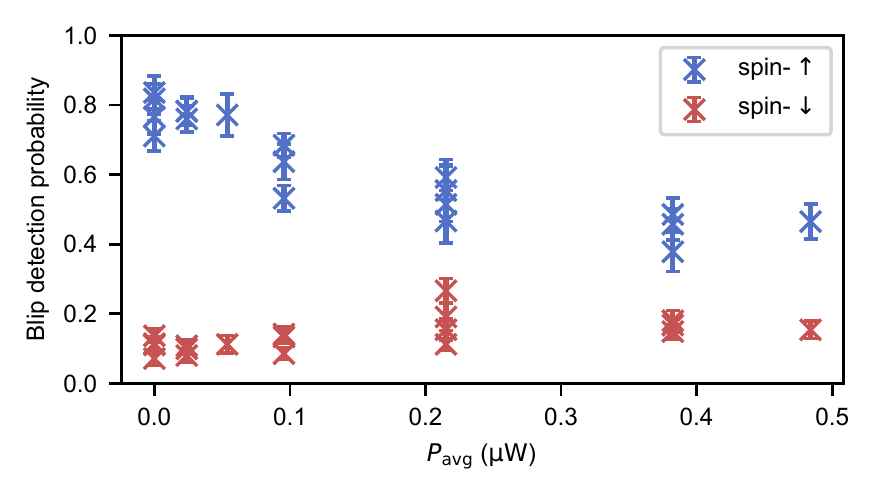}
	\caption{\label{fig:supfig3} \textbf{Effect of RF pulses on electron spin readout | }
	The probability to observe a quantum dot to reservoir transition for an electron spin readout is affected by the average applied RF power ($P_\mathrm{avg} = P_\mathrm{pulse}\times t_\mathrm{pulse}/t_\mathrm{rep}$, with $P_\mathrm{pulse}, t_\mathrm{pulse}$ the RF pulse power and duration and $t_\mathrm{rep}$ the time between successive RF pulses). Powers are referred to the output of the RF source. Probabilities are evaluated by fitting repeated ESR frequency scans. For the nuclear-electron entanglement experiment presented in Fig.~3e, average powers of 0.003 to 0.008  $\mathrm{\mu W}$ were applied.}
\end{figure}

\clearpage
\bibliographystyle{naturemag}